\documentclass[11pt,onecolumn,amssymb,nofootinbib]{revtex4}
\usepackage{amsmath, amsthm, amscd, amssymb}
\usepackage{bm}
\usepackage{bbm}

\begin{document}

\title{\bf Collinearity constraints for on-shell massless particle three-point functions, and implications for
allowed-forbidden $n+1$-point functions}

\author{Stephen L. Adler}
\email{adler@ias.edu} \affiliation{Institute for Advanced Study,
Einstein Drive, Princeton, NJ 08540, USA.}

\begin{abstract}
A simple collinearity argument implies that
the massless particle three-point function of helicities $h_1, h_2, h_3$  with corresponding real-valued four-momenta $k_1, k_2, k_3$ taken
as all incoming or all outgoing (i.e., $k_1 +k_2 +k_3=0$), vanishes by helicity conservation unless $h_1+h_2+h_3=0$.   When
any one particle with four-momentum $k$ is off mass shell, this constraint no longer applies; a  forbidden amplitude
with $h_1+h_2+h_3\neq 0$ on-shell can be nonzero off-shell, but vanishes proportionally to $k^2$ as $k$ approaches mass shell.  When an on-shell
forbidden amplitude is coupled to an allowed $n$-point amplitude to form an $n+1$ point function, this  $k^2$ factor in the forbidden amplitude cancels the $k^2$  in the propagator, leading to a $n+1$-point function that has no pole at $k^2=0$.  We relate our results for
real-valued four-momenta to the corresponding selection rules that have been derived in the on-shell literature for complexified four-momenta.

\end{abstract}

\maketitle

A number of recent papers \cite{mcgady}, \cite{benincasa}, \cite{schuster}, \cite{britto} have studied the properties of on-mass-shell three- and four-point functions for massless particles by employing factorization and pole counting constraints on the four-point $S$-matrix, together with complex continuation of four-momenta.  Our purpose in
this note is to show that strong constraints governing on-mass-shell three-point functions for massless particles with real-valued
four-momenta can be obtained by using
a collinearity argument that appeared in the context of high energy neutrino reactions \cite{adler1} and photon splitting in a constant
external magnetic field \cite{adler2}.

Consider the amplitude ${\cal A}(k_1 h_1k_2 h_2|k_3 h_3)$ for a particle of four-momentum $k_1$ and helicity $h_1$ combining with a particle
of four-momentum $k_2$ and helicity $h_2$ to give an outgoing particle of four-momentum $k_3$ and helicity $h_3$. We take all four-momenta
to be real-valued rather than complexified as in  \cite{mcgady}, \cite{benincasa}, \cite{schuster}, \cite{britto}.    Since all particles
are propagating forward in time, their energies are non-negative,
\begin{equation}\label{positivity}
k_1^0=|\vec k_1|~,~~~
k_2^0=|\vec k_2|~,~~~
k_3^0=|\vec k_3|~~~,
\end{equation}
and the on-mass-shell condition states that
\begin{equation}\label{mass_shell}
k_1^2=k_2^2=k_3^2=0~~~.
\end{equation}
We assume that the amplitude $\cal{A}$ depends on no variables other than the ones explicitly shown.  Squaring the four-momentum conservation
condition $k_3=k_1+k_2$ and using Eqs. \eqref{positivity} and \eqref{mass_shell} gives
\begin{equation}\label{square}
0=k_3^2=k_1^2+k_2^2+2k_1 \cdot k_2 =2k_1 \cdot k_2 =2|\vec k_1| |\vec k_2| \big(1-\cos\theta\big)~~~,
\end{equation}
with $\theta$ the angle between the three-vector momenta $\vec k_1$ and $\vec k_2$.  Equation \eqref{square} implies
that $\cos\theta=1$, that is, the three-vectors $\vec k_1$ and $\vec k_2$, and hence also $\vec k_3$, are collinear.
Rotational invariance around the common direction of propagation then implies the helicity conservation
condition
\begin{equation}\label{helicity1}
h_1+h_2=h_3~~~.
\end{equation}
When the four-momentum $k_3$ is taken as incoming, so that $k_1+k_2+k_3=0$, its energy is non-positive and its helicity is reversed in
sign, in agreement with the standard convention \cite{elvang} used in the spinor helicity formalism.  Equation \eqref{helicity1} then becomes
\begin{equation}\label{helicity2}
h_1+h_2+h_3=0~~~;
\end{equation}
clearly the same constraint also holds if all three particles are taken as outgoing.
Thus amplitudes ${\cal A}(k_1h_1k_2h_2k_3h_3)\equiv{\cal A}(k_1 h_1k_2 h_2|-k_3 -h_3)$ with  $ h_1 +h_2 +h_3 \neq 0$  must vanish. This implies, for example, that the amplitude for a massless spin $\frac{3}{2}$ particle to absorb
a real photon must vanish, since $\pm\frac{3}{2} \pm 1$ can never give $\pm \frac{3}{2}$.

The constraint that we have derived no longer holds when any one of the three particles is off-shell.  If the off-shell four-momentum is
denoted by $k$, an amplitude that is forbidden on-shell by the helicity constraint is no longer forbidden, but develops a kinematic zero
proportional to $k^2$.  Consider now an amplitude in which an allowed $n$-point function $B(...,kh)$, which can involve massless
or massive particles with momenta and helicities denoted by ..., is linked by exchange of a massless particle with four-momentum $k$ and helicity $h$ to a helicity-forbidden three-point function
$A(k h  k_3 h_3 k_4 h_4)$  with $h+h_3+h_4 \neq 0$.  The  corresponding  $n+1$-point amplitude is proportional to
\begin{equation}\label{fourpoint}
 \sum_h B(...,kh) \frac{1}{k^2} A(k h  k_3 h_3 k_4 h_4)~~~.
\end{equation}
But since the helicity-forbidden amplitude on the right has a kinematic zero,
\begin{equation}\label{zero}
A(k h  k_3 h_3 k_4 h_4)=k^2 R(k h  k_3 h_3 k_4 h_4)
\end{equation}
with $R$ regular as $k$ approaches mass shell, Eq. \eqref{fourpoint} reduces
to the form
\begin{equation}\label{fourpoint1}
 \sum_h B(...,kh)  R(k h  k_3 h_3 k_4 h_4)~~~.
\end{equation}
This has no pole at $k^2=0$ in the variable $k^2$, and so is part of the background analytic in $k^2$  that is not determined by
polology arguments.  This scenario applies to the scattering of a charged spin $\frac{3}{2}$ particle in a non-constant electric or
magnetic field, and so the amplitude \cite{adler3} for this scattering need not vanish kinematically.

We conclude by comparing our results with the on-shell rules for massless particle three-point functions obtained by McGady and Rodina \cite{mcgady}.
They give a range of cases for which three-point functions with $h_1+h_2+h_3\neq 0$ are non-vanishing for complexified
four-momenta.  Our results show that in the limit of real four-momenta, all of these amplitudes must vanish.  Conversely,
McGady and Rodina show that all $h_1+h_2+h_3=0$ amplitudes with complexified momenta vanish except for the case $0+0+0=0$
corresponding to the three scalar meson coupling of $\phi^3$ theory.  Taking the real four-momentum limit of their result, this
shows that all helicity-allowed amplitudes other than the $\phi^3$ vertex vanish for on-shell massless particles.  For the
cases of a massless spin-1/2 particle scattering off a massless spin 0 or spin 1 particle to another massless spin-1/2
particle, this can be verified directly from the Feynman rules for the vertex.  For a massless Dirac particle, the Lorentz spinor
$u(k,h)$ is given by
\begin{equation}\label{dirac}
u(k,h)= N(k)\left(
\begin{array}{c}
1_2\\
\vec \sigma \cdot \hat k\\
\end{array}
\right)\chi(h)
\end{equation}
with $N(k)$ a normalization constant, $\hat k=\vec k/|\vec k|$ a unit vector, $\vec \sigma$ the Pauli matrices, $1_2$ a $2 \times 2$
unit matrix,
and $\chi(h)$ a 2 component spinor carrying the helicity information.  From this formula, together with the
Dirac gamma matrix
$\gamma^0 =\rm{diag}(1,-1)$, one immediately sees
that $\overline {u}(k_1,h_1) u(k_2,h_2)=0$ in the collinear case $\hat k_1=\hat k_2$, which shows
vanishing of the vertex in which a massless spin-1/2 particle absorbs a  massless spin-0 particle.  Similarly, one sees that
$\overline {u}(k_1,h_1) \,\vec \gamma \cdot \vec e \,u(k_2,h_2)=0$ when $\vec k_1=\vec k_2$ and $\vec k_1 \cdot \vec e=0$,
which shows vanishing of the vertex in which a massless spin-1/2 particle absorbs a massless spin-1 particle with transverse
polarization vector $\vec e$.
In both cases, the contribution from the ``small'' components of the Dirac spinor exactly cancels the contribution from
the ``large'' components of the Dirac spinor, something that does not happen when the spinor describes a massive spin-1/2
particle.

The author wishes to thank E. Witten for raising the question addressed in this note,  L. Rodina and D. McGady for informative
conversations and email correspondence, and G. Paz for comments on the initial arXiv version.

\end{document}